\documentclass[runningheads,a4paper]{llncs}

\usepackage{amssymb}
\usepackage{graphicx}
\usepackage{subfigure}
\usepackage{caption}
\usepackage{multirow}
\usepackage{booktabs}
\usepackage{array}

\usepackage{url}
\urldef{\mailsc}\path|yasir_alf@yahoo.com|    
\newcommand{\keywords}[1]{\par\addvspace\baselineskip
\noindent\keywordname\enspace\ignorespaces#1}

\begin{document}

\mainmatter  

\title{Power Consumption Analysis of a Modern Smartphone}

\titlerunning{Lecture Notes in Computer Science: Authors' Instructions}

%
%
\author{Muhammad Yasir Malik}
\authorrunning{Power Consumption Analysis of a Modern Smartphone}

\institute{Seoul National University}

%
%

\maketitle

\begin{abstract}
This paper presents observations about power consumption of a 
latest smartphone. Modern smartphones are powerful devices with 
different choices of data connections and other functional modes.
This paper provides analysis of power utilization for 
these different operation modes. Also, we present power consumption by 
vital operating system (OS) components.

\end{abstract}

\section{Introduction}

Smartphones saw their advent some 16 years ago with the introduction of Nokia Communicator series. 
Growth of smartphones only started increasing exponentially in recent years. According to Strategy Analytics, 
total number of smartphones in the world has reached to one billion. It is expected to hit two billion in three years.

This shows how much part smartphones are playing, and will play, in our daily lives. Latest smartphones are handy, 
powerful and are equipped with various functionalities. 

\subsection{Power in Smartphones}
With these advantages, some constraints come along in the 
form of power consumption. It is normal to recharge (or replace) smartphone batteries at least once or more times 
a day. Similarly, smartphone battery appears to drain over night, when smartphone is partially or fully idle.

All smartphone companies are taking this issue seriously, as we can notice bigger batteries with newer models of smartphones. 
Extra features like flight mode, automatic and manual power save modes are also included in smartphones to save energy.

Modern smartphones are capable of using more than one data networks.
Also, smartphones have certain features and modes that claim to save 
power consumption thus increasing the battery usage hours.

\subsection{Our Contributions}

In this report, we will try to analyze power consumption of a modern 
smartphone while it uses LTE and WiFi data network connections. 
Moreover, power consumption of the smartphone in some important 
modes, namely flight and power-save modes, are computed.
 
Along with these observations, power utilization of smartphone for 
normal usage and while it is in sleep mode is also measured.
 
We believe this paper will provide a baseline for further analysis 
in the works related to power consumption in modern smartphones.

\section{Device Under Test (DUT)}

We used SHV-E120S, also known as Samsung Galaxy S2 HD LTE, for our tests. Platform, display and battery specifications of the tested 
phone are given below.

\subsection{DUT Specifications}

Following table lists some notable specifications of our DUT.
\begin{table}
\centering
    \caption{Specifications of DUT}
    \begin{tabular}{ | l | l | p{8cm} |}
    \hline
    Network & 2G Network &   GSM 850 / 900 / 1800 / 1900  \\ \cline{2-3}
            & 3G Network &    HSDPA 850 / 900 / 1900 / 2100, \\
  	        &            &    HSDPA 900 / 2100               \\ \cline{2-3}
            & 4G Network & 	  LTE 800 / 1800 / 2600  \\ \hline
    Data & Speed & LTE, HSDPA, 21 Mbps \\ 
            & WLAN  & Wi-Fi 802.11 a/b/g/n, DLNA, Wi-Fi Direct, Wi-Fi hotspot  \\ \cline{2-3}
            & Others & GRPS, EDGE, BluetoothWednesday  \\  \hline
    Display & Screen & Super AMOLED capacitive touchscreen, 16M colors   \\ \cline{2-3}
            & Size   & 720 x 1280 pixels, 4.65 inches \\ \hline
    Battery & Capacity & Li-Ion 1850 mAh \\ \cline{2-3}
            & Stand-by & Up to 320 h (2G) / Up to 290 h (3G) \\ \cline{2-3}   
            & Talk time &	Up to 12 h 40 min (2G) / Up to 5 h 50 min (3G) \\ \hline
   Platform &    OS 	& Android OS, v2.3 (Gingerbread) \\ \cline{2-3}
            & Chipset 	& Qualcomm MSM8660 Snapdragon \\ \cline{2-3}
            & CPU 	    & Dual-core 1.5 GHz Scorpion  \\ \cline{2-3}
            & GPU 	    & Adreno 220  \\ \cline{2-3}
            & Sensors 	& Accelerometer, gyro, proximity, compass \\ \hline
   Others   & Messaging &	SMS(threaded view), MMS, Email, Push Mail, IM, RSS \\ \cline{2-3}
            & Browser 	&   HTML, Adobe Flash \\ \cline{2-3}  
            & Radio 	&  No  \\ \cline{2-3}
            & GPS 	    & Yes, with A-GPS support \\ \hline        
    \end{tabular}
    \end{table}

\subsection{Operating System Specifications}

Factors related to power consumption are not limited to hardware alone; operating systems (OS) components also 
contribute to battery drainage.

In this section, we briefly describe some OS fragments that seem to affect power more than other OS parts.

\begin{enumerate}
  \item \textbf{Android System} \\
  Android system performs most of functions of OS, and consumes more power than other components of OS. 
  Android system includes, more importantly, service manager, sensor manager,  com.android.settings, 
  tvoutserver, dmbserver etc.  \\
  
  \begin{enumerate}
    \item Service manager  \\
    Android runtime uses the ServiceManager to add services, and to find them, but the ServiceManager 
    itself is accessed via a Binder. It manages, lists and add all service that are running in the system. 
    A Service is an application component representing either an application's desire to perform a 
    longer-running operation while not interacting with the user or to supply functionality for other 
    applications to use.
    
    \item Sensor manager  \\
    Modern day smartphones have many interesting sensors available to enhance user experience, by 
    providing novel applications. Sensors provide accurate information about device positioning, 
    orientations and environmental conditions. 
    
    Sensor manager is responsible for managing the operations of these sensors. Some of the sensors 
    available in our device are:
    
    \begin{enumerate}
    \item Accelerometer  \\
     Measures the acceleration force in $m/s^2$ that is applied to a device on all three physical 
     axes (x, y, and z), including the force of gravity.
    \item Gyroscope  \\
     Measures a device's rate of rotation in $rad/s$ around each of the three physical axes (x, y, and z).
    \item Magnetic field  \\
     Measures the ambient geomagnetic field.
    \item Orientation  \\     
     Measures degrees of rotation that a device makes around all three physical axes (x, y, z).
\end{enumerate}
    \item Others    \\
    Other notable parts of OS are Tvoutserver, DMBserver, com.android.settings and 
    com.android.fatorysettings. \\
  \end{enumerate}
  
  \item \textbf{OS Kernel}    \\
  Kernel in OS links facilitates the communication between hardware and software layers. 
  Kernel includes drivers, scripts, security modules and firmware to name a few. Kernel keeps 
  consuming power regardless of whether they are operating at any moment or not. \\

  \item   \textbf{Android core apps}    \\
  They are basically applications which are installed by default with Android OS. 
  Email service i.e. Gmail, keyboard, Internet Voice calling, clock, calendar and 
  alarm etc. are some of the core apps available in Gingerbread 2.3. \\

  \item   \textbf{Microbes}     \\
  Microbes is live wallpaper application available in Gingerbread 2.3. \\

  \item	\textbf{Mediaserver}     \\
  Mediaserver in Android accesses gallery, audio and video files available on device and plays them.   \\

  \item	\textbf{Goolge sevices framework}    \\
  Google services framework allows the device to communicate with Google for various purposes 
  i.e. application licensing, AdMob ads, In-App billing. Firm upgrade, accessing Google cloud 
  and Google maps are also handled with this framework.   \\

  \item	\textbf{Anti-virus}    \\
   Anti-virus program may run on smartphone at all times to ensure security.\\
   
  \item	\textbf{Misc. services}    \\
   Other important Android, Google or stand-alone services running on devices are email 
   synchronization program, message, store, dictionary etc.  \\
   
  \item	\textbf{Other common apps}    \\
   Most commonly used applications such as Facebook, Twitter, Skype and mobile messaging 
   service Kakaotalk were installed on DUT. Android OS comes with some already installed applications
   i.e. Youtube, Google Store, Maps, Navigation to name a few.
\end{enumerate}

\subsection{Test Equipment}

We used power monitor by Monsoon Solutions Inc. for our measurements. 
Exact output voltage can be determined by considering the resistances, 
internal resistance of test equipment and source channel resistance.
\begin{center}
$V_{exact} = V_{output} – (I_{drawn} \times R)$
\end{center}
Normally, the resistance of 20 gauge wire is 0.012 $\Omega$/ft. The length of 
cable used for testing was about two feet.
According to above equation, setting the output voltage $V_{out}$ at 3.76 V provides us 3.7 V as 
real output voltage. 
Specifications including the accuracy of the current for USB
channel are given in Table 2.
\begin{table}
  \centering
    \caption{USB channel specs of test equipment}
    \begin{tabular}{ | c | >{\centering}p{4cm} | c |}
    \hline
  Component &   Min &   Max  \\ \hline
  Input voltage range   & 2.1 V  &  5.4 V \\ \hline  
  Continuous current   & -   &  1.0 A   \\ \hline  
  Fine current scale - range &  -  &   40 mA   \\ \hline
  Fine current scale - resolution  &  2.86 uA  & - \\ \hline
Fine current scale - accuracy  &  +/- 1\% or +/- 50 uA (whichever is greater) & -  \\ \hline
Coarse current scale - range  &  30 mA &  4.5 A \\ \hline
Coarse current scale - resolution  & 286 uA &  -  \\ \hline
Coarse current scale - accuracy &  +/- 1\% or +/- 1 mA (whichever is greater) & -\\ \hline
VBUS capacitance to GND & 22uF +/- 20\%  & -\\ \hline
    \end{tabular}
    \end{table}

\section{Modes of Operation}

We will compare power performance of DUT under various scenarios. In this section, 
we describe some of these situations.

\begin{enumerate}

\item \textbf{Data Networks}    \\
LTE and Wi-Fi data networks are both used during experimentation. We analyze both networks 
during their \textit{normal} and \textit{sleep} mode. \\

\item \textbf{Normal Operation}       \\
During normal operation, users use their smartphones in a typical way. 
Different users utilize their smartphones differently. Though based on average current
consumption, some optimized criteria can be set. Thus, we took many samples in order to 
make certain the lower and upper bounds of power consumption. Average of these samples is 
presented in this work.\\

\item \textbf{Sleep}              \\
Sleep mode is inactive mode of \textit{normal operation}. Users perform no activity of 
any kind with their smartphones. It is equivalent to the passive mode of smartphones at night.
We started calculations after initial power surges subsided. \\

\item \textbf{Flight Mode}       \\
Generally flight mode in phones disables data connection i.e. no data enters or leaves the phone.
Only native applications can be used in this mode.   \\

\item \textbf{Power Save Mode}    \\
Latest smartphones include this mode to enhance the battery life by taking steps needed to
decrease the current consumption. We will also analyze the effect of this mode in practical 
in next section.
\end{enumerate}           

Table 3. depicts functionalities of smartphone which are activated or de-activated 
during different modes of DUT.  
\begin{table}[htbp]
  \centering
  \caption{Functions operating in different modes}
    \begin{tabular}{|c|c|c|c|c|c|c|c|c|}    \hline
    \multirow{2}{*}{Activity} & \multicolumn{2}{c}{LTE} & \multicolumn{2}{|c}{Wi-Fi} & \multicolumn{2}{|c|}{Flight mode} & \multicolumn{2}{c|}{Power-save mode} \\ \cline{2-9}
          & Normal & Sleep & Normal & Sleep & Normal & Sleep & Normal & Sleep \\ \hline
    Network (LTE or Wi-Fi) & Yes   & Yes   & Yes   & Yes   & No    & No    & Yes   & Yes \\ \hline
    GPS   & Yes   & Yes   & Yes   & Yes   & No    & No    & No    & No \\ \hline
    Bluetooth & Yes   & Yes   & Yes   & Yes   & No    & No    & No    & No \\ \hline
    Notifications & Yes   & Yes   & Yes   & Yes   & No    & No    & Yes   & Yes \\ \hline
    Email synchronization & Yes   & Yes   & Yes   & Yes   & No    & No    & No    & No \\ \hline
    Apps running in background & Yes   & Yes   & Yes   & Yes   & No    & No    & Yes   & Yes \\ \hline
    CPU power optimization & No    & No    & No    & No    & No    & No    & Yes   & Yes \\ \hline
    Brightness intensity optimization & No    & No    & No    & No    & No    & No    & Yes   & Yes \\ \hline 
    \end{tabular}%
  \label{tab:addlabel}%
\end{table}%
\section{Observations}

In this section, we present results and analysis of power consumption
for the scenarios we defined in the previous section.
Results are based on more than 7.5 million samples collected during the testing.
Table 4. lists power consumed under different conditions. 
\begin{table}[htbp]
  \centering
  \caption{ Power consumption analysis for different modes}
    \begin{tabular}{|c|c|c|c|c|c|c|c|c|}    \hline
    \multirow{2}{*}{Quantity} & \multicolumn{2}{c}{LTE} & \multicolumn{2}{|c}{Wi-Fi} & \multicolumn{2}{|c|}{Flight mode} & \multicolumn{2}{c|}{Power-save mode} \\ \cline{2-9}
          & Normal & Sleep & Normal & Sleep & Normal & Sleep & Normal & Sleep \\ \hline
 Consumed energy (μAh) & 238543.19 & 15251.97 & 200477 & 5460.7& 228526 & 5460.7 & 276512.64 & 32529.06 \\ \hline
 Average power (mW) & 2200  & 133.09 & 1748.85 & 366.28& 1993.35 & 47.65 & 2411.8 & 283.75\\ \hline
 Average current (mA) & 587.87& 35.42 & 465.59 & 97.5  & 530.73 & 12.68 & 642.18 & 75.54\\ \hline
 Battery life (hrs) & 3.58 & 59.29 & 4.51 & 21.54 & 3.96 & 165.59 & 3.27 & 27.8\\ \hline
\end{tabular}%
  \label{tab:addlabel}%
\end{table}%
Here, the observations about the power consumption 
are presented in graphical form for ease of analysis.

At sleep mode, WiFi network seems to be consuming more energy than others.
Flight mode is more enery efficient. In normal operation, however, WiFi 
shows better performance in regard to energy consumed.
Power-save mode does not help in saving energy, as shown in Figure 1.  

In Figure 2. average power consumption shows the same behavior as that of consumed energy.
Wifi network shows better power performance at sleep mode, but perform 
oppositely in normal functioning. Similarly, power-save mode does not shows better 
performance.

In Figure 3., average current for sleep and normal mode depicts 
the same obseravtions as depicted in previous figures.
Average current consumption for LTE network is far lower that WiFi
and power-save mode. This is true for both sleep and normal mode.

In Figure 4., based on the analysis, smartphone in flight mode will save more power thus
extending the battery life.
LTE networks show better performance than WiFi network and power-save mode in sleep.

\begin{figure}
\centering
 \includegraphics[height=6cm,width=14cm]{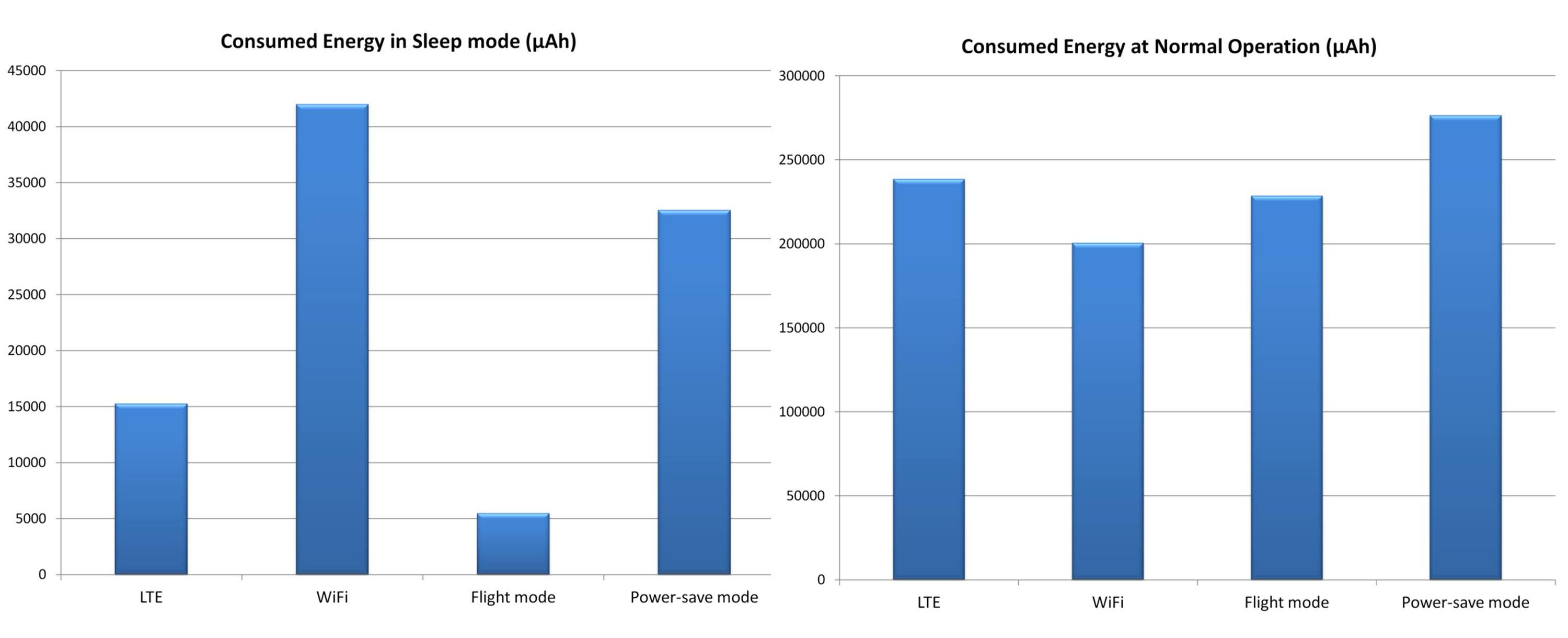}
\caption{Consumed energy in Sleep and Normal mode} \label{Figure 1.}
\end{figure}

\begin{figure}
\centering
 \includegraphics[height=6cm,width=14cm]{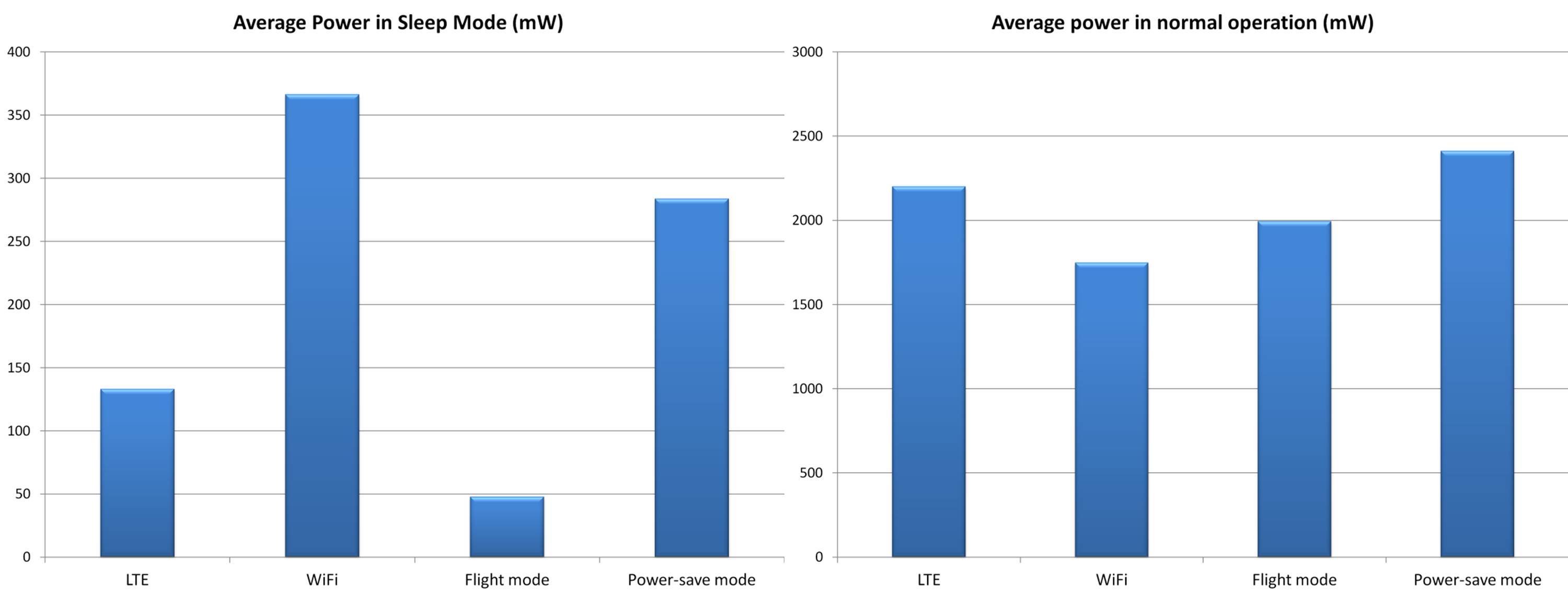}
\caption{Average power in Sleep and Normal mode} \label{Figure 2.}
\end{figure}

\begin{figure}
\centering
 \includegraphics[height=6cm,width=14cm]{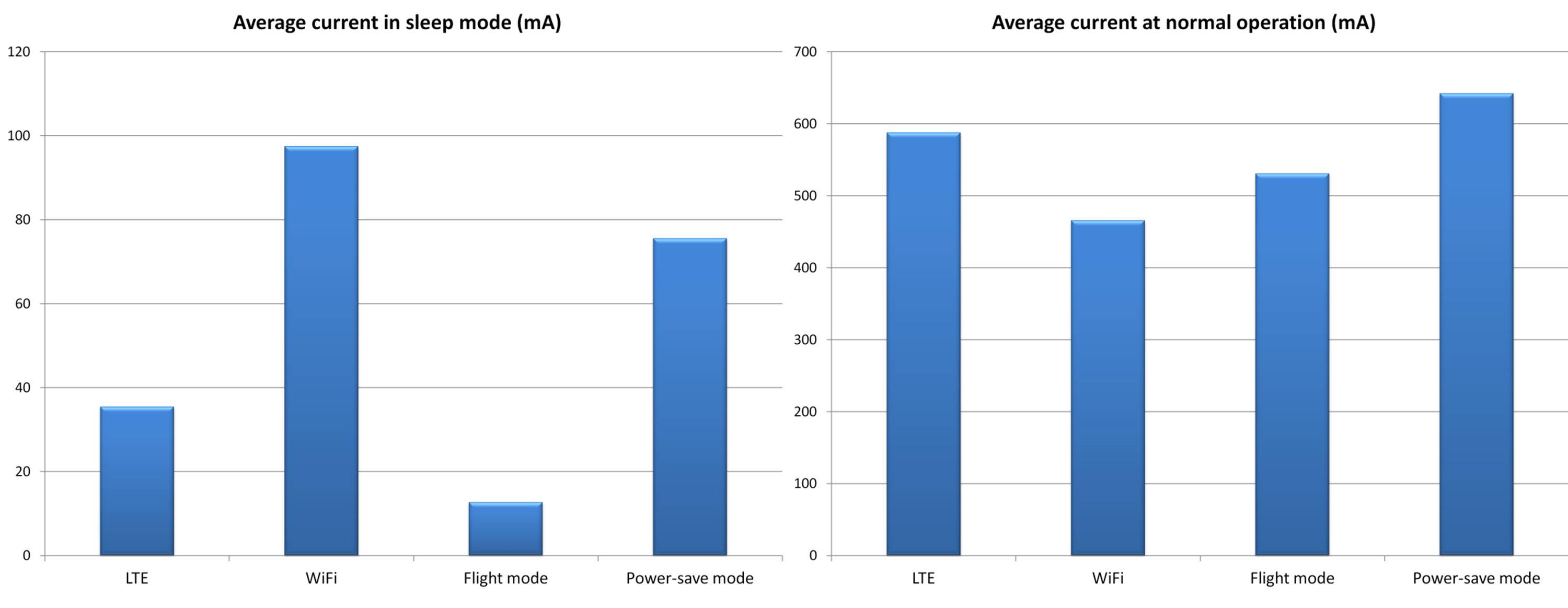}
\caption{Average current in Sleep and Normal mode} \label{Figure 3.}
\end{figure}

\begin{figure}
\centering
 \includegraphics[height=6cm,width=14cm]{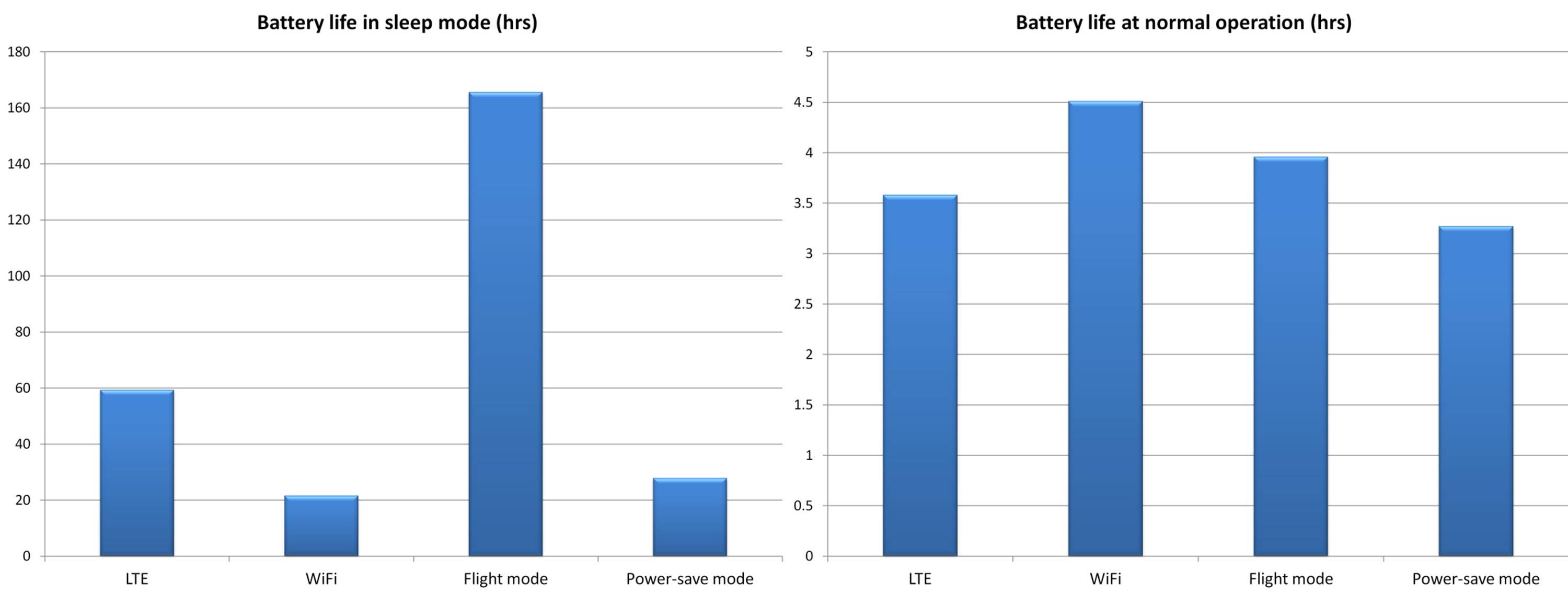}
\caption{Battery life in Sleep and Normal mode} \label{Figure 4.}
\end{figure}

\newpage

Table 5. shows the share of important Android components, defined briefly in Section 3,
in overall power consumption.
\begin{table}[htbp]
  \centering
  \caption{Functions operating in different modes}
    \begin{tabular}{|c|c|c|c|c|c|c|c|c|}
    \hline
    \multirow{2}{*}{Activity} & \multicolumn{2}{c}{LTE} & \multicolumn{2}{|c}{Wi-Fi} & \multicolumn{2}{|c|}{Flight mode} & \multicolumn{2}{c|}{Power-save mode} \\ \cline{2-9}
          & Normal & Sleep & Normal & Sleep & Normal & Sleep & Normal & Sleep \\ \hline
    Android system level & 18.4   & 24.4   & 20   & 19.9   & 8.7    & 28.3    & 21.0   & 21.8 \\ \hline
    Kernel   & 4.2   & 6.4   & 4.0   & 4.2   & 1.4    & 6.8    & 3.5   & 5.2 \\ \hline
    Android core apps & 2.6   & 4.9   & 2.1   & 2.3   & 58.1    & 4.4    & 3.3   & 2.9 \\ \hline
    Google services & -     & 1.5   & 0.9   & 1.3   & 0.3    & 2.3    & 1.3   & 3.8 \\ \hline
    Anti-virus program & 3.1   & 6.8   & 9.6   & 26.2  & 12.2   & 9.0    & 13.8   & 18.9 \\ \hline
    Facebook & 4.6   & 3.5   & 4.2   & 0.9   & 0.6    & 4.1    & 10.1   & 3.7 \\ \hline
    Internet & 11.6  & -    & 23.7   & -    & -      & -      & 6.3   & - \\ \hline
    Microbes & 3.0   & 4.0   & 3.2   & 7.2   & 1.5    & 8.1    & 2.5   & 9.0 \\ \hline
    Kakaotalk & 0.7   & 1.2   & 2.1   & 9.8   & -    & 3.9    & 2.4   & 4.8 \\ \hline
    Google maps & 0.3   & 1.1   & 0.9   & 2.6   & 0.9    & 1.8    &  0.9   & 1.4 \\ \hline
    Dialer & 2.6   & 1.5   & 0.6   & 0.8   & -    & 1.4   &  0.8   & 1.3 \\ \hline
    System mediaserver & -   & -   & 1.0   & -   & 0.3    & -   & 0.7   & 0.4 \\ \hline
    Google Play & 0.2   & -   & -   & 0.5   & -    & 0.8   & 0.4   & 0.8 \\ \hline
    Media & -   & -   & -   & -   & -    & -   & 0.4   & 0.5 \\ \hline
    Email and messages & -   & 1.1   & -   & -   & -    & 0.4   & -   & - \\ \hline
    \end{tabular}%
  \label{tab:addlabel}%
\end{table}%

Figure 5. shows the power consumed in normal mode, whereas Figure 6. shows the power
consumed in sleep mode. 
\begin{figure}
\centering
 \includegraphics[height=5cm,width=12cm]{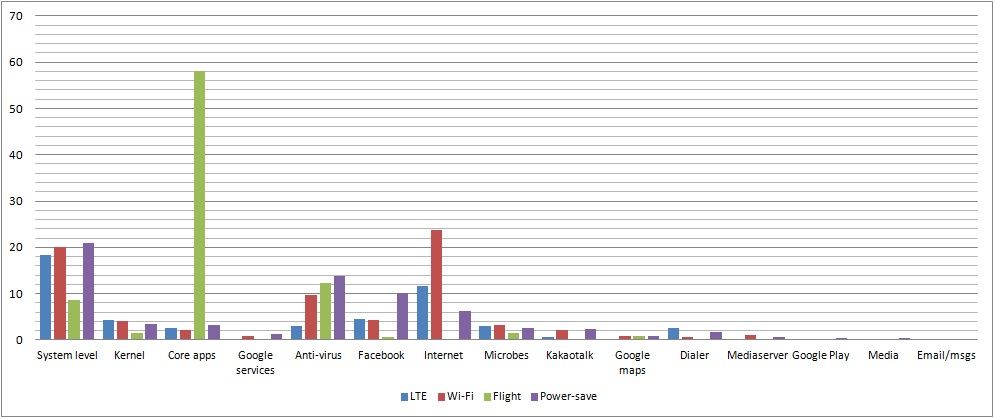}
\caption{Consumed energy in Normal mode} \label{Figure 5.}
\end{figure}

\begin{figure}
\centering
 \includegraphics[height=5cm,width=12cm]{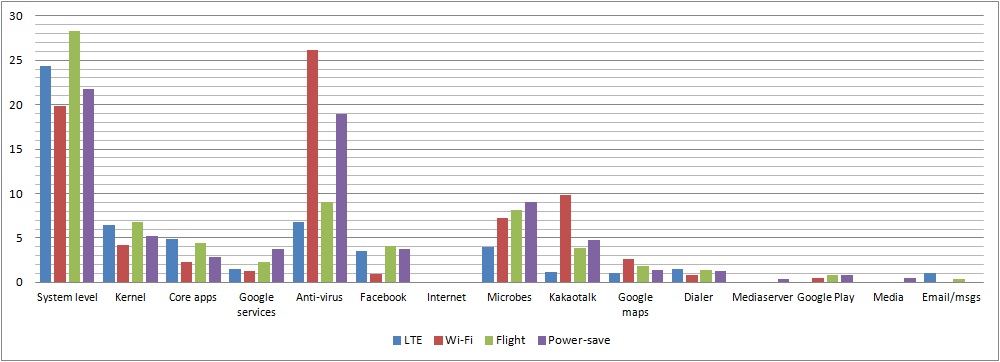}
\caption{Consumed energy in Sleep mode} \label{Figure 6.}
\end{figure}

\newpage

\section{Conclusion}
In this paper, we anaylzed power consumption of a latest smartphone for 
different working modes. This study showed that some modes that are meant for
saving power i.e. power-save mode, are not that efficient in saving power.
In sleep, putting our smartphones in flight mode can help us in saving most
of its energy, and thus extend its battery consumption.
LTE, although a new network, performs better in sleep mode. In normal operation, however,
it is still expensive for power.

We also anaylzed OS-level power consumption of a latest smartphone for 
different working modes. This work may form the basis for further in-depth
analysis on power consumption in smartphones. Furthermore, power optimization can also benefit from
this work.

\end{document}